\documentclass[reprint, groupedaddress]{revtex4-2}

\usepackage{tensor}
\usepackage{amsmath}
\usepackage{amssymb}
\usepackage{enumerate}
\usepackage{mathrsfs}
\usepackage{mathtools}
\usepackage{graphicx}
\usepackage{stmaryrd}
\usepackage{tikz-cd}
\usepackage{bbm}
\usepackage{comment}
\usepackage[compat=1.1.0]{tikz-feynman}
\usepackage{twistor}
\usepackage{graphicx}
\usepackage{dcolumn}
\usepackage{bm}
\usepackage[hidelinks]{hyperref}

\newcommand{\mbb}{\mathbb}
\newcommand{\msf}[1]{\mathsf{#1}}

\newcommand{\op}{\operatorname}

\newcommand{\WZW}{\mathrm{WZW}}
\newcommand{\til}{\tilde} 
\renewcommand{\d}{\mathrm{d}}
\renewcommand{\c}{\sf{c}} 
\renewcommand{\b}{\sf{b}} 

\newcommand{\vac}[1]{\left|#1\right\rangle}

\begin{document}


\title{Top-down holography in an asymptotically flat spacetime}

\author{Kevin Costello}
\affiliation{Perimeter Institute for Theoretical Physics, Waterloo, ON, CA}
\email{kcostello@perimeterinstitute.ca}

\author{Natalie M. Paquette}
\affiliation{Department of Physics, University of Washington, Seattle, WA, USA}
\email{npaquett@uw.edu}

\author{Atul Sharma}
\affiliation{The Mathematical Institute, University of Oxford, UK}
\email{atul.sharma@maths.ox.ac.uk}

\date{\today}

\begin{abstract}
We propose a holographic duality for a four-dimensional WZW model with target manifold $\SO(8)$, coupled to scalar-flat K\"ahler gravity on an asymptotically flat, four dimensional background known as the Burns metric. The holographic dual is a two dimensional chiral algebra built out of gauged beta-gamma systems with $\SO(8)$ flavor. We test the duality by matching two-point correlators of soft gluon currents with two-point gluon amplitudes, and their leading OPE coefficients with collinear limits of three-point gluon amplitudes. 
\end{abstract}

\maketitle


\section{Introduction}

Holography on asymptotically flat spacetimes has been a long-standing open problem, with existing gauge-gravity dualities relating matrix quantum mechanics to flat space string and M-theory (e.g. \cite{BFSS, Polchinski, Susskind, BMN, IKKT}). In this note, we describe a new kind of holographic duality wherein the bulk is a four-dimensional asymptotically flat, self-dual spacetime. The duality relates certain models of self-dual gauge and gravitational theories in the bulk to a two-dimensional chiral algebra living on a Riemann sphere.

This example arises from a generalization of twisted holography \cite{CG, CP1}, whereby the B-model topological string \footnote{These examples often work in the supergravity limit of the B-model, known as Kodaira-Spencer theory \cite{BCOV, CosLiBCOV}.} on a ``backreacted'', i.e. complex structure-deformed geometry is dual to a B-brane worldvolume theory. The latter is typically a chiral algebra, supported on a complex submanifold of the undeformed flat space geometry. In these twisted examples, remarkably, the backreaction on the closed string side of the duality truncates to a polynomial in $N$. This renders small-$N$ computations around the original flat background equivalent, up to normalization of chiral algebra operators, to holographic $1/N$-expansions (at least for observables that do not scale with $N$). Moreover,  open-closed duality in the small-$N$ expansion has recently been formalized mathematically as Koszul duality \cite{Costello:2017fbo, CP1, PW}, which permits efficient computation of the boundary chiral algebra OPEs.

The present example is rooted in recent developments connecting the celestial holography program to twisted holography and twistor theory \cite{Costello:2022wso,Bu:2022dis}. In flat space, it was observed that the physics of soft gluons and gravitons is governed by certain 2d chiral algebras. These chiral algebras often fail to be associative when one incorporates quantum effects \cite{Costello:2022upu}, which may complicate the ultimate formulation of dual CFT constructions \footnote{See also \cite{Ball:2021tmb}, and see \cite{Ren:2022sws} for the highly stringent constraints associativity places on 4d EFT coefficients.}. If, however, the 4d theory can be uplifted to a local holomorphic theory on twistor space \cite{Costello:2021bah}, an associative chiral algebra, \textit{including quantum effects corresponding to higher-loop collinear amplitudes}, is guaranteed. The $G=\SO(8)$ WZW$_4$ theory is one of these distinguished theories that can therefore serve as a simple toy model, and its quantum consistency when coupled to the gravitational sector is guaranteed by a Green-Schwarz anomaly cancellation mechanism in the type I topological string \cite{Costello:2019jsy} on twistor space. (The type I topological string may be best understood as a subsector of twisted, Omega-deformed type IIB string theory.) \newline
\-\hspace{1em}The resulting twisted open-closed string theory is our undeformed bulk theory, and we systematically incorporate backreaction, per the standard holographic setup, by wrapping $N$ D1-branes on the zero-section of the twistor fibration $\simeq \mathbb{CP}^1$, the Riemann sphere. This deforms the spacetime geometry from flat space to the asymptotically flat \textit{Burns space} \cite{Lebrun91explicitself-dual}. The D1-branes, as usual, support the dual, boundary chiral algebra. The usual identification of the twistor $\mathbb{CP}^1$ with the celestial sphere at null infinity then makes this a concrete, top-down toy model of celestial holography as envisioned by e.g. \cite{Strominger:2017zoo,Pasterski:2021raf} and references therein.
We will elaborate on the string theory uplift of this example, and present additional details and checks of our proposed duality, in a forthcoming companion paper \cite{CPS}.\newline
\-\hspace{1em}In the remainder of this note we introduce our bulk gauge/gravitational theory and the dual chiral algebra, describe the holographic dictionary between bulk states and chiral algebra operators, and provide explicit checks of the duality in 2 and 3-point computations. The supplementary materials contain computational details for the chiral algebra OPEs as well as sample computations of bulk 2- and 3-gluon amplitudes to leading nontrivial order.


\section{Bulk theory}

Our bulk spacetime is self-dual (i.e.\ it has a self-dual Weyl tensor), so it is most conveniently formulated in Euclidean signature. Let $x^\mu\in\R^4$ and introduce double-null coordinates on $\R^4$ \cite{Mason:1991rf} 
\be
x^{\al\dal} \vcentcolon= \frac{1}{\sqrt2}\begin{pmatrix}x^0+\im x^3&&x^2+\im x^1\\-x^2+\im x^1&&x^0-\im x^3\end{pmatrix}
\ee
where $\al=1,2$, $\dal=\dot1,\dot2$ are Weyl spinor indices. In terms of these, we can set up complex coordinates on $\R^4\simeq\C^2$,
\be
u^{\dal} \vcentcolon= x^{1\dal}\,,\qquad \hat u^{\dal} \vcentcolon= x^{2\dal} = (-\overline{u^{\dot2}},\overline{u^{\dot1}})\,.
\ee
Also let $||u||^2 \vcentcolon= |u^{\dot1}|^2+|u^{\dot2}|^2=\frac{1}{2}\,\delta_{\mu\nu}x^\mu x^\nu$, and denote by $\p=\d u^{\dal}\,\p/\p u^{\dal}$, $\dbar = \d\hat u^{\dal}\,\p/\p \hat u^{\dal}$ the holomorphic and anti-holomorphic exterior derivatives on $\C^2$. 

We equip $\C^2-\{0\}$ with the Riemannian metric
\be\label{burns}
\d s^2 = ||\d u||^2 + \frac{|u^{\dot1}\d u^{\dot2}-u^{\dot2}\d u^{\dot1}|^2}{||u||^4}\,.
\ee
This is a self-dual metric known as the \emph{Burns metric} \cite{burns1986twistors, Lebrun91explicitself-dual, Dunajski:2019smi}. It is asymptotically flat, in the sense that $g_{\mu\nu} = \delta_{\mu\nu} + O(||u||^{-2})$ at large $||u||$ \cite{lebrun1988counter}. It has zero scalar curvature but is \emph{not} Ricci flat. Its isometry group is the $\U(2)$ rotating the dotted index on $u^{\dal}$.  
Moreover, it is K\"ahler with K\"ahler potential
\be\label{kahler}
K = ||u||^2 + \log||u||^2\,.
\ee
Let $\widetilde\C^2=\{(u^{\dal},[\zeta^{\dal}])\in\C^2\times\CP^1\,|\,u^{\dot1}\zeta^{\dot2}=u^{\dot2}\zeta^{\dot1}\}$ denote the blowup of $\C^2$ at the origin. Away from $u^{\dal}=0$, it is diffeomorphic to $\C^2-\{0\}$, whereas the origin $u^{\dal}=0$ is replaced by a copy of $\CP^1$ called the exceptional divisor. The metric \eqref{burns} extends smoothly to $\widetilde\C^2$. In fact, \eqref{burns} is most naturally obtained by restricting the direct sum of the flat and stereographic metrics on $\C^2\times\CP^1$ to $\widetilde\C^2$ \cite{lebrun1988poon}.

We will refer to $\widetilde\C^2$ equipped with the Burns metric as \emph{Burns space}. In the past, it has featured in the spacetime foam models of \cite{Hawking:1979pi}, being conformally diffeomorphic to (an affine patch of) $\CP^2$ equipped with its Fubini-Study metric via the inversion map $u^{\dal}\mapsto u^{\dal}/||u||^2$. It has also been conjectured to result from the backreaction of D1-branes in topological string theory on twistor space \cite{Hartnoll:2004rv}, a fact that we will prove in the companion paper \footnote{The $\text{K\"ahler}$ potential perturbation associated to a single closed-string field emanating from a D1 brane was computed in \cite{Costello:2021bah} to be $\log||u||^2$. What is surprising is that this already satisfies the equations of motion and that there are no further corrections.}. 


The field theory we study on Burns space is the $\WZW_4$ model \cite{Donaldson:1985zz, Nair:1991bf, Losev:1995cr,Costello:2021bah} for $\op{SO}(8)$. This is a $\sigma$-model governing maps $g(u,\hat u)$ from Burns space to the group manifold of $\op{SO}(8)$, with action
\begin{multline}\label{eq:WZW}
\frac{N}{8\pi^2}\int_{\widetilde\C^2} \p\dbar K\wedge\op{tr}\,(g^{-1}\p g \wedge g^{-1}\dbar g)\\
- \frac{N}{24\pi^2}\int_{\widetilde\C^2\times[0,1]}\p\dbar K\wedge\tr\,(\tilde g^{-1}\d\tilde g)^3
\end{multline}
where $N>0$ is a 4d analogue of the Kac-Moody level, $\tilde g$ is an extension of $g$ to $\widetilde\C^2\times[0,1]$ satisfying the boundary conditions $\tilde g|_{\widetilde\C^2\times\{0\}}=\mathbbm{1}$, $\tilde g|_{\widetilde\C^2\times\{1\}}=g$, and $\d\tilde g$ represents its exterior derivative on $\widetilde\C^2\times[0,1]$. 
The first term in this action is the standard kinetic term for a $\sigma$-model, and the second is a 5d Wess-Zumino term. 

It is crucial that the path integral of WZW$_4$ be independent of the choice of 5d extension. With our normalization, this requires that $(\im N/2\pi)\,\p\dbar K$ be a representative of integral cohomology $H^2(\widetilde\C^2,\Z)$ \cite{Losev:1995cr}. In particular, integrating it over the exceptional divisor of $\widetilde\C^2$ yields
\be
\int_{\CP^1}\frac{\im N}{2\pi}\,\p\dbar K = N\,.
\ee
So, we demand the quantization condition $N\in\Z_+$ \footnote{This is the 4d analogue of the quantization of Kac-Moody level in 2d WZW models.}.
 
Classically, the WZW$_4$ model provides a gauge-fixed formulation of self-dual Yang-Mills \cite{Yang:1977zf,Mason:1991rf}. The Yang-Mills gauge field is taken to be $A=-\dbar g\,g^{-1}$, 
and the equation of motion $\p\dbar K\wedge\dbar(g^{-1}\p g)=0$ following from \eqref{eq:WZW} -- known as Yang's equation -- guarantees that this satisfies the self-dual Yang-Mills equation. This match with gauge theory does not persist at loop level. The model does appear to be related to the 4d $\cN=2$ heterotic string \cite{Ooguri:1991ie}, although the precise connection has yet to be clarified.
 
We couple this model to a certain gravitational field \cite{Costello:2021bah}, which is a dynamical scalar field interpreted as a perturbation of the K\"ahler potential. Its action governs K\"ahler manifolds with vanishing scalar curvature. 
These may have non-vanishing Ricci curvature, and so the gravitational theory does not relate to Einstein gravity in the same way as WZW$_4$ relates to self-dual Yang-Mills. Nevertheless, the coupled system admits an anomaly free uplift to the type I open+closed topological B-model on twistor space \cite{Costello:2021bah,Bittleston:2020hfv,Penna:2020uky}, to which one can apply twisted holography \cite{CG, CP1}. We refer to our gravitational theory as \emph{scalar-flat K\"ahler gravity}. It will not play a major role in the rest of this note.
 
The main result of this work is a holographic duality matching the collinear singularities in scattering amplitudes of $\WZW_4$ on Burns space, coupled to the dynamical K{\"a}hler scalar, with OPE coefficients of the large $N$ limit -- the \emph{planar limit} -- of a family of 2d chiral algebras (with defects).  In this note, we present explicit checks of the duality in the planar limit for open string operators in the chiral algebra, which correspond to tree-level Yang-Mills collinear singularities; additional planar computations matching the gravitational degrees of freedom will appear in \cite{CPS}.
A similar chiral algebra is also known to control the collinear singularities of Yang-Mills amplitudes on flat space, at tree level \cite{Guevara:2021abz, Ball:2021tmb} and sometimes at loop level \cite{Costello:2022wso,Costello:2022upu}.  

Compared to known results about celestial holography, our proposal has an important new feature.   We have a description of the dual chiral algebra \emph{non-perturbatively}, i.e. at finite bulk coupling $1/\sqrt N$.  This is in stark contrast to previous results, where the chiral algebra is described explicitly only at tree level or one loop. We conjecture that this strong form of the duality, analogous to the strong form of the AdS/CFT correspondence, holds.  

$\WZW_4$ is a non-renormalizable quantum field theory, placed on a curved manifold.  It is perhaps surprising that our duality gives an exact (conjectural) description of collinear limits in this theory at finite coupling. This should be compared to known results about two-dimensional integrable quantum field theory, where exact non-perturbative results about scattering amplitudes can be derived even for models which are non-renormalizable \cite{Fateev:1992tk}.


\section{The holographic dual}

The dual theory is a 2d chiral algebra living on $\CP^1$. In \cite{CPS}, we will arrive at it as the worldvolume theory of D1-branes wrapping a projective line in the twistor space of $\R^4$. It is given by the BRST reduction of a collection of symplectic bosons by the gauge group $\operatorname{Sp}(N)$ (with the conventions that the fundamental representation of $\operatorname{Sp}(N)$ is the symplectic vector space of dimension $2N$). Additionally, it has matter content consisting of spin $\frac{1}{2}$ fields valued in symplectic representations of $\operatorname{Sp}(N) \times \op{SO}(8)$, where $\op{SO}(8)$ is a flavor symmetry.

There are two kinds of fields (arising from 1-5 and 1-1 strings \footnote{From the point of view of a 10d type I string compactified on twistor space, the fields arise from 3-7 and 3-3 strings. When placed in an $\Omega$-background transverse to twistor space, the D3 and D7-branes become Euclidean D1 and D5 branes} respectively):
\begin{align}
    I = I_{i m} &\in \C^8 \otimes \C^{2 N}\,,\\
    X = X_{\dot\alpha m n} &\in \C^2 \otimes \sideset{}{_{\substack{\text{trace}\\\text{free}}}^2}\bigwedge \C^{2N}.
\end{align}
Here $m$ is an index in $\C^{2N}$, $i$ in the fundamental of $\op{SO}(8)$, and $\dot\alpha$ in the fundamental of $\SU(2)$, which acts by isometries of Burns space. The tensor $X$ is trace-free: $X_{\dot\alpha m n} \omega^{mn} = 0$, with $\omega^{mn}$ the symplectic form on Sp$(N)$.  

The OPEs are 
\begin{align}
I_{i m}(z_1) I_{jn}(z_2) &\sim \frac{\delta_{ij}\,\omega_{mn}}{z_{12}}\,,\\
X_{\dot\alpha mn}(z_1) X_{\dot\beta rs}(z_2) &\sim \frac{\eps_{\dot\alpha \dot\beta}}{z_{12}} \left(\omega_{m[r|}\omega_{n|s]}-\frac{\omega_{mn}\omega_{rs}}{2N}\right) 
\end{align}
where $z_{ij}\equiv z_i-z_j$, $\eps_{\dal\dot\beta}$ is the 2-dimensional Levi-Civita symbol, and $\omega_{mn}$ is the inverse of $\omega^{mn}$. To this algebra, we adjoin $\b$, $\c$ ghosts living in the adjoint of $\op{Sp}(N)$, with the usual BRST operator.  The BRST cohomology in the large $N$ limit is our conjecture for the holographic dual chiral algebra.

These fields can be written down for any $\op{SO}(k)$ flavour symmetry group. Only when $k = 8$ does the BRST operator square to zero.  This is the CFT counterpart of the fact \cite{Costello:2019jsy, Costello:2021bah} that the type I topological string is anomaly free only for $\op{SO}(8)$.

It is worth remarking that this algebra is the chiral algebra associated by the construction of \cite{Beem:2013sza} to a well-known family of 4d $\mathcal{N}=2$ SCFTs, with $\op{SO}(8)$ flavor symmetry, $\op{Sp}(N)$ gauge symmetry and matter as above. In the case $N=1$ this theory is $\SU(2)$ gauge theory with $N_f = 4$, as studied by Seiberg and Witten \cite{SW}.

At large $N$, the BRST cohomology can be readily computed using the method explained in \cite{CG}, based on classic results of homological algebra (this computation is done in detail in \cite{MoosavianZhou}). To write down the operators, we use $\omega^{mn}$ to raise an index so that $X_{\dot1 m}{}^n$, $X_{\dot2 m}{}^n$ are $2 N \times 2 N$ matrices. By raising and lowering $\op{Sp}(N)$ indices we can treat $I_{im}$ as being a collection of $8$ vectors or covectors.  

To describe the open string operators, let us introduce an $\SU(2)$ doublet $\til{\lambda}_{\dot\alpha}$.  A basis for open string operators are the coefficients of $\til{\lambda}_{\dot1}{}^k \til{\lambda}_{\dot2}{}^l$ in the generating function
\begin{equation}
	\begin{split} 
		J_{ij} [\til{\lambda}] (z) &=  I_{i} \exp \Bigl((2N)^{-\frac{1}{2}} \eps^{\dot\alpha \dot\beta} \til{\lambda}_{\dot\alpha} X_{\dot\beta}  \Bigr)  I_j \label{eqn_cft_state} \\	
		&= \sum_{k,l=0}^\infty \frac{1}{k!\, l!}\;\til{\lambda}_{\dot1}{}^k\, \til{\lambda}_{\dot2}{}^l\, J_{ij}[k,l](z). 
	\end{split}	
\end{equation}
This is in the adjoint of $\SO(8)$, so we will freely replace composite indices like $ij$ with adjoint indices $a,b,c,\dots$ in what follows. Similarly, the terms with $k+l = n$ are in the spin $n/2$ representation of $\SU(2)$. 

There are also two towers of closed-string states, which will not play an important role in this note:
\begin{equation*}
\begin{split}
	E[\til{\lambda}](z) &= \op{Tr}\bigl( \exp ((2N)^{-\frac{1}{2}}\eps^{\dot\alpha \dot\beta} \til{\lambda}_{\dot\alpha} X_{\dot\beta}) \bigr) \\ 
	F[\til{\lambda}](z)  &= \op{Tr}\bigl(\eps^{\dot\alpha \dot\beta} X_{\dot\alpha} \partial X_{\dot\beta} \, \exp ((2N)^{-\frac{1}{2}} \eps^{\dot\gamma\dot\delta} \til{\lambda}_{\dot\gamma} X_{\dot\delta} ) \bigr) 
+ \cdots \end{split}
\end{equation*}
where the ellipses in $F[\til\lambda]$ indicate terms involving ghosts.

In the planar limit the OPEs can be computed by elementary Wick contractions:
\begin{align}
    &J_a[\til{\lambda}_1] (z_1)\, J_b[\til{\lambda}_2] (z_2)\sim \frac{f_{ab}{}^c}{z_{12}}\,J_c[\til\lambda_1+\til\lambda_2](z_2)  \nonumber\\ 
	&- \frac{[1\,2]\,f_{ab}{}^c } {z_{12}^2}  \int_0^1 \d \omega_1 \int_0^1\d \omega_2\; J_c[\omega_1 \til{ \lambda}_1 + \omega_2\til{\lambda}_2 ](z_2)   \nonumber	\\ 
	&+\cdots - 2N\sum_{n=0}^\infty \frac{(-[1\,2])^n} { n!\,n!} \frac{\kappa_{ab} \, \mathbf{1}}{z_{12}^{n+2}} \label{eqn_cft_ope}
\end{align}
where $\kappa_{ab} = \delta_{il}\delta_{jk}-\delta_{ik}\delta_{jl}$ in terms of composite indices $a=ij,b=kl$, and we used the standard conventions $[1\,2] = \eps^{\dal\dot\beta}\til\lambda_{1\dot\beta}\til\lambda_{2\dal}$ and $\eps^{\dot\gamma\dal}\eps_{\dot\gamma\dot\beta} = \delta^{\dal}{}_{\dot\beta}$ for Weyl spinor contractions. The ellipsis denotes terms involving three or more contractions, whereas the contribution of the identity arises from maximal contractions. The negative level of $-2N$ indicates that our celestial holographic dual is non-unitary. Some minor subtleties of this computation are also discussed in the supplementary material.

\subsection{Defects and boundary conditions}

We will relate OPEs of this CFT to amplitudes on Burns space. In the remainder of this note, by abuse of language, we will refer to our (holomorphic, non-unitary) chiral algebra as a 2d CFT; we emphasize that, per the usual twist of \cite{Beem:2013sza}, this is in fact a subsector of an ordinary unitary CFT in \textit{four} spacetime dimensions.  

The isometries of Burns space are $\U(2)$, so that the complexification $\GL(2,\C)$ is a symmetry of the analytically continued scattering amplitudes.  This is in addition to the $\SO(8)$ symmetry.   As written, however, the CFT has an $\SL(2,\C) \times \SL(2,\C) \times \SO(8)$ symmetry, where one $\SL(2,\C)$ is a flavor symmetry rotating the dotted index on $X_{\dal}$ and the other is 2d conformal symmetry. 

The more precise statement of our dictionary involves a CFT with defects at $0$ and $\infty$, which break the $\SL(2,\C)$ conformal symmetry to $\C^\times$.  The defects are modules generated by vacuum vectors $\vac{v_0}$ and $\vac{v_\infty}$   annihilated by certain modes of $J_a[k,l]$:
\begin{equation}
\begin{split}
\oint \d z\, z^n J_a[k,l] \vac{v_0} &= 0 \text { for } n \ge k+l+1 \\
\oint \d z\, z^n J_a[k,l] \vac{v_\infty} &= 0 \text { for } n \le -1\,. \\
\end{split}
\end{equation}
We will expand on the geometric and stringy origin of these defects in the companion paper. They have also been studied from the perspective of half-BPS surface defects in 4d $\cN=2$ theories in \cite{BRP,Cordova:2017mhb}.



\section{States in the bulk theory}

To study gluon perturbation theory, we parametrize the WZW$_4$ field as $g = \e^{\phi}$, where $\phi$ is an adjoint valued scalar field. As its extension, we take $\tilde g = \e^{t\phi}$ for $t\in[0,1]$. The action \eqref{eq:WZW} then expands to
\be\label{phiac}
\int_{\widetilde\C^2}\frac{N\,\p\dbar K}{8\pi^2}\wedge\tr\biggl(\p\phi\wedge\dbar\phi - \frac{1}{3}\,\phi\,[\p\phi,\dbar\phi]+O(\phi^4)\biggr)\,.
\ee
The linearized field equation of $\phi$ obtained from this action is simply the Laplace equation on Burns space,
\be\label{laplace}
\biggl(\eps^{\dal\dot\beta} + \frac{ u^{\dal}\hat u^{\dot\beta}}{||u||^4}\biggr)\,\frac{\p^2\phi}{\p u^{\dal}\p\hat u^{\dot\beta}}=0\,.
\ee
As in flat space, we find a family of analytic solutions of \eqref{laplace} labeled by a complex null momentum $p_{\al\dal}=\lambda_\al\til\lambda_{\dal}$, where $\til\lambda_{\dal}\in\C^2$ whereas $\lambda_\al=(z^{-1},1)$ for some $z\in\C^\times$:
\be\label{burnsmom}
    \phi_a(z, \til\lambda) = \frac{\msf{T}_a\,\e^{\im p\cdot x/2}}{z}\left(\cos\frac{\psi\, p\cdot x}{2}+\frac{\im}{\psi}\,\sin\frac{\psi\, p\cdot x}{2}\right)
\ee
(suppressing $x^{\al\dal}$ dependence on the left). In this expression, $\msf{T}_a$ is a generator of the gauge algebra $\mathfrak{so}(8)$ and
\begin{align}
    &p\cdot x = \frac{[u\,\til\lambda]}{z} + [\hat u\,\til\lambda]\,,\\
    &\psi = \sqrt{1+\frac{1}{z}\,\frac{[u\,\til\lambda][\hat u\,\til\lambda]}{||u||^2}\frac{4}{(p\cdot x)^2}}\,.
\end{align}
Here, $[u\,\til\lambda] = u^{\dal}\til\lambda_{\dal}$ and $[\hat u\,\til\lambda]=\hat u^{\dal}\til\lambda_{\dal}$ are spinor contractions. 

The state \eqref{burnsmom} extends to the blowup $\widetilde\C^2$. It also admits a series expansion
\be\label{largeu}
\frac{\msf{T}_a}{z}\sum_{j=0}^\infty\biggl(-\frac{1}{z}\frac{[u\,\tilde\lambda][\hat u\,\tilde\lambda]}{||u||^2}\biggr)^j\frac{{}_1F_1(j+1,2j+1\,|\,\im p\cdot x)}{(2j)!}
\ee
whose leading $j=0$ term is the flat space momentum eigenstate $\e^{\im p\cdot x}$, up to the important normalization factor of $1/z$. This formally acts as a ``large $||u||$'' asymptotic expansion.  

We can also Taylor expand this state in $\til{\lambda}_{\dal}$ to write:
\begin{equation}
\phi_a(z,\tilde{\lambda}) = \sum_{k,l=0}^\infty \frac{1}{k!\, l!}\; \til{\lambda}_{\dot1}{}^k\, \til{\lambda}_{\dot2}{}^l\, \phi_a[k,l](z) \,.
\end{equation}
We will refer to $\phi_a[k,l](z)$ as soft modes, as they are analogous to the coefficients in the flat space soft expansion of $\e^{\im p\cdot x}$. We propose the holographic dictionary
\begin{equation}
\phi_a[k,l](z)\; \longleftrightarrow\; J_a[k,l](z) 
\end{equation}
and so 
\begin{equation} \label{eq:dictionary}
    \phi_a(z, \til{\lambda} )\; \longleftrightarrow \; J_a[ \til{\lambda}] (z) .
\end{equation}

\section{Conformal blocks}

Our duality identifies the WZW$_4$ model, coupled to scalar-flat K\"ahler gravity, with the 2d (defect) CFT described above. One aspect of proving a global holographic correspondence would be to prove that scattering amplitudes on Burns space match CFT correlators, e.g.\ in the planar limit. The presence of defects introduces a subtlety in making this statement. The CFT with defects has an infinite dimensional space of conformal blocks, making the definition of correlators ambiguous.

A conformal block is a way to define the correlators of local operators compatible with the poles that arise from OPEs, and the poles when operators are brought to the defects. It is easy to see that there are many ways to define correlators subject to these constraints.  E.g., the most general two-point function of the Kac-Moody current $j_a = J_a[0,0]$ compatible with the poles is 
\be\label{cblock}
	\la v_0|\, j_a(z_1)\, j_b(z_2)\,|v_\infty\ra = \kappa_{ab}\left( -\frac{2N}{z_{12}^2} +  \frac{C}{z_1 z_2}\right)
\ee
for some arbitrary constant $C$. Both terms in \eqref{cblock} are symmetric in $1,2$ and transform in the same way under the defect conformal group action $z_i\mapsto sz_i$, $s\in\C^\times$. 

For the purposes of this note, we may state our holographic correspondence as a match between states and their OPEs on both sides. On the bulk side, OPEs are collinear limits in scattering amplitudes. 


In practical terms, however, we can account for the conformal block ambiguity as follows.  The CFT correlators (in any conformal block) are polynomials in $(z_i-z_j)^{-1}$ and in $z_i^{-1}$.  The conformal block ambiguity is contained in expressions depending on $z_i^{-1}$. We expect a match between CFT correlators and WZW$_4$ amplitudes on Burns space, \emph{up to terms depending on the $z_i^{-1}$}.  


\section{Tests of the duality}

\subsection{Two-point functions}

Amplitudes in Euclidean signature can be defined and computed via the on-shell effective action \cite{Arefeva:1974jv, Jevicki:1987ax}, and then analytically continued to more physical signatures if necessary. To test our duality, we begin by computing the two point amplitude of WZW$_4$.  

Let $\phi_i\equiv\phi_{a_i}(z_i,\til\lambda_{i\dal})$, where $i$ is a particle label. The 2-point tree amplitude of WZW$_4$ is given by the symmetrized on-shell kinetic term,
\be
A(1,2) = \frac{N}{8\pi^2}\int_{\widetilde\C^2}\p\dbar K\wedge\tr\!\left(\p\phi_{1}\wedge\dbar\phi_{2}\right) + (1\leftrightarrow2)\,.
\ee
Calculating this on Burns space by plugging in the expansion \eqref{largeu} for $\phi_1,\phi_2$ reveals a delightfully simple result:
\be\label{2ptamp}
A(1,2) = -\frac{N}{z_{12}^2}\;J_0\Biggl(2\,\sqrt{\frac{[1\,2]}{z_{12}}}\Biggr)\;\tr(\msf{T}_{a_1}\msf{T}_{a_2})
\ee
with $J_0$ denoting a Bessel function of the first kind. Up to the color factor, this can be shown to match the 2-point amplitude of a conformally coupled scalar on $\CP^2$ independently computed in \cite{Hawking:1979pi}.

Under our proposed identification of $J_a[\til{\lambda}](z)$ with $\phi_a (z,\til{\lambda})$, if we normalize the generators so that $\tr(\msf{T}_a\msf{T}_b) = 2\,\kappa_{ab}$, then this exactly matches with the coefficient of the identity operator in the OPE of $J_a[\til{\lambda}](z)$.  Indeed, we can resum the coefficient of the identity operator $\mathbf{1}$ in \eqref{eqn_cft_ope} to get
\begin{equation} 
	J_{a}[\til{\lambda}_1] (z_1)\,  J_{b}[\til{\lambda}_2] (z_2) \sim -\frac{2N\kappa_{ab}\mathbf{1}}{z_{12}^2}\; J_0 \Biggl(2\, \sqrt{\frac{ [1\,2]} {z_{12} } } \Biggr)\,.  
\end{equation}
To give a flavor of the computation, we derive $\eqref{2ptamp}$ at $O(z_{12}^{-2})$ in the supplementary materials, with more details to appear in \cite{CPS}.


\subsection{OPE coefficients}

Define the `holographic OPE' $\phi_1\cdot\phi_2$ between two states $\phi_1,\phi_2$ by demanding that the linear combination
\begin{equation} 
	\varepsilon_1 \phi_1 + \varepsilon_2 \phi_2 + \varepsilon_1 \varepsilon_2 \,\phi_1 \cdot \phi_2  
\end{equation}
satisfy the non-linear equations of motion of \eqref{phiac} modulo $\varepsilon_i^2$. Then the 2-point amplitude of the (off-shell) state $\phi_1\cdot\phi_2$ with a third state $\phi_3$ gives rise to the 3-point amplitude of $\phi_1,\phi_2,\phi_3$ (after symmetrization over $1,2,3$). This is closely related to the perturbiner approach to scattering amplitudes \cite{Rosly:1996vr,Rosly:1997ap}. 


From the equation of motion of \eqref{phiac}, the OPE satisfies
\begin{equation}\label{eq:OPE}
\p\dbar K \wedge\biggl(\p\dbar (\phi_1 \cdot \phi_2) + \frac{[\p\phi_1,\dbar\phi_2]+[\p\phi_2,\dbar\phi_1]}{2}\biggr)=0\,,
\end{equation} 
having used $\p\dbar K\wedge\p\dbar\phi_i=0$ to drop $\p\dbar K\wedge[\p\dbar\phi_1,\phi_2]$, etc. The solution of \eqref{eq:OPE} for $\phi_1\cdot\phi_2$ is found to be
\begin{multline}\label{opesol}
-\int_{v\in\til\C^2}G(u,v)\;\p\dbar K\wedge([\p\phi_1,\dbar\phi_2]+[\p\phi_2,\dbar\phi_1])\Bigr|_v    
\end{multline}
where $G(u,v)$ for $u^{\dal},v^{\dal}\in\C^2-\{0\}$ is the massless scalar propagator on Burns space
\be\label{burnsprop}
G(u,v) = -\frac{1}{8\pi^2}\,\frac{1}{||u-v||^2+||u||^{-2}||v||^{-2}|[u\,v]|^2}\,,
\ee
and $[u\,v] = \eps_{\dal\dot\beta}u^{\dot\beta}v^{\dal}$ as usual. \eqref{burnsprop} can be derived by a conformal rescaling of the propagator of a conformally coupled scalar on $\CP^2$ given in \cite{Hawking:1979hw}.

For example, to compute $\phi_1\cdot\phi_2$ at zeroth order around flat space, set $\phi_i=\msf{T}_{a_i}\,\e^{\im p_i\cdot x}/z_i$, $\p\dbar K = \eps_{\dal\dot\beta}\,\d u^{\dal}\wedge\d\hat u^{\dot\beta}$, and $G(u,v)=-1/8\pi^2||u-v||^2$. Evaluating \eqref{opesol} yields
\begin{align}
\phi_1 \cdot \phi_2 &= \frac{[\msf{T}_{a_1},\msf{T}_{a_2}]}{z_{12}}\,\frac{z_1+z_2}{2z_1z_2}\;\e^{\im(p_1+p_2)\cdot x}\nonumber \\  &\sim \frac{[\msf{T}_{a_1},\msf{T}_{a_2}]}{z_{12}}\,\frac{\e^{\im x^{\al\dal}\lambda_{2\al}(\tilde\lambda_1+\tilde\lambda_2)_{\dal}}}{z_2} + O(z_{12}^0)
\end{align}
where recall that $\lambda_{i\al}=(z_i^{-1},1)$. It is easily checked that this data solves \eqref{eq:OPE}. This takes the form of the standard Kac-Moody OPE at order $1/z_{12}$ if we choose the normalization convention $[\msf{T}_{a_1},\msf{T}_{a_2}] = f_{a_1a_2}{}^c\,\msf{T}_c$ for the structure constants.

More generally, we have evaluated \eqref{opesol} on Burns space to first order in $[1\,2]$ to find
\begin{multline} \label{opematch}
	\phi_1 \cdot \phi_2 \sim 
	  \frac{f_{a_1 a_2}{}^c}{z_{12}}\; \phi_c( z_2, \til{\lambda}_1 + \til{\lambda}_2) \\
	 -  \frac{[1\,2]\,f_{a_1 a_2}{}^c}{z_{12}^2 } \int_0^1\d\omega_1\int_0^1\d\omega_2 \;  \phi_c( z_2, \omega_1 \til{\lambda}_1 + \omega_2 \til{\lambda}_2)\\ + O([1\,2]^2)\,. 
\end{multline}
While computing this in the OPE limit, we have dropped terms regular in $z_{12}$. The result \eqref{opematch} matches the chiral algebra OPE displayed in \eqref{eqn_cft_ope} under the dictionary \eqref{eq:dictionary}.


Lastly, one can also directly compute the 3-gluon amplitude by plugging the $\phi_i$ into the cubic interaction vertex of \eqref{phiac}. At zeroth order in the square brackets $[1\,2],[2\,3],[3\,1]$, we find a current algebra 3-point function
\be
\frac{-2Nf_{a_1a_2}{}^c\,\kappa_{ca_3}}{z_{12}z_{13}z_{23}}
\ee
which enables us to directly probe the $j_a(z_1)\,j_b(z_2)$ OPE. We derive this result in the supplementary material.


\section{Discussion}
We have presented a top-down toy model of holography in an asymptotically flat spacetime, motivated by constructions from celestial holography, twisted holography and twistor theory. Following the standard AdS/CFT dictionary, we have presented explicit 2 and 3-point checks relating chiral algebra OPEs with asymptotic bulk observables in the Burns metric. As is familiar from the AdS case, these computations did not rely on the presence of dynamical gravity in the bulk. Indeed, although our chiral algebra has a local stress tensor operator (which should be contrasted with the celestial algebras in \cite{Costello:2022wso}), the closed string deformations of our model only capture fluctuations of the K{\"a}hler potential. This is a consequence of the defects at the antipodes of $\mathbb{CP}^1$, which partially break diffeomorphism invariance. We will present checks of the duality involving the closed string sector in \cite{CPS}. 

A related conceptual difference between our toy model and standard (untwisted) holography is the infinite dimensional space of conformal blocks associated to the chiral algebra. On the bulk side of the duality, different choices of conformal block correspond to different ways of filling in the asymptotic boundary of Burns space with a bulk system.  We can achieve this by using the Burns geometry, but with  local operators integrated over the $\CP^1$ at the core of Burns space. There are infinitely many local operators, and so infinitely many ways to do this; we plan to study this aspect of the dictionary in future work. 

In standard holographic dualities, this issue does not arise (at least in the planar limit) because diffeomorphisms are gauged in the bulk, which prevents the insertion of local operators.  The ``gravitational'' theory in our model is a gauge-fixed theory of a dynamical K\"ahler potential, where diffeomorphisms have \textit{not} been gauged. Clearly it would be desirable to have a toy model in which one could directly access features of dynamical (quantum) gravity like bulk reconstruction, black hole transitions, spacetime entanglement \& quantum information, etc. Likewise, since our twistorial theory is highly non-generic, it will be important to better understand which features may be relaxed (and how), and which break down, in more realistic examples.

Nevertheless, we believe our toy model, with its twisted string origins, is a concrete and illustrative starting point for asymptotically flat holography in string theory. Twisted holography is also sensitive to certain nonperturbative effects in $N$ \cite{BG}, and it will be enlightening to formalize the map between $O(N)$ effects in the bulk and boundary (gauge theory instantons/Skyrmions, bulk giant gravitons, \!\ldots), as well as $O(N^2)$ effects, perhaps by studying the chiral algebra on other boundary geometries (see also \cite{Atanasov:2021oyu}) and/or finding connections to Kleinian black holes analogous to those in \cite{Crawley:2021auj}.

One would also like to find a similar top-down realization of asymptotically flat holography for self-dual Einstein gravity. The study of anomaly free twistor uplifts of models of self-dual general relativity was recently initiated in \cite{Bittleston:2022nfr}. On the dual side, two-dimensional chiral defect CFTs computing graviton amplitudes in self-dual vacuum spacetimes have already emerged as a bridge between Einstein gravity, hyperk\"ahler geometry and celestial holography \cite{Adamo:2021bej,Adamo:2021lrv,Adamo:2022mev}. Burns space too can be viewed as an Einstein-Maxwell instanton \cite{Dunajski:2006vs} instead of as a vacuum solution of scalar-flat K\"ahler gravity. It would be very interesting to understand whether it could emerge from brane backreaction in a twistor string uplift of self-dual general relativity coupled to self-dual gauge theory. 

We also expect the techniques of this paper to generalize to self-dual gravitational backgrounds in $(2,2)$-signature, where the study of scattering amplitudes has received considerably more attention. Similarly, other curved backgrounds relevant to celestial holography have been discussed in \cite{Pasterski:2020pdk,Fan:2022vbz,Casali:2022fro,Gonzo:2022tjm}, and it will be worthwhile trying to find their uplifts to possible string theory backgrounds. 

\smallskip

\textit{Acknowledgments:} We would like to thank Tim Adamo, Chris Beem, Roland Bittleston, Eduardo Casali, Adam Chalabi, Maciej Dunajski, Joel Fine, Lionel Mason, Sujay Nair and David Skinner for useful conversations. We are also grateful to Tim Adamo, Lionel Mason, Sabrina Pasterski, David Skinner, and Andrew Strominger for comments on the draft. NP is supported by the University of Washington and the DOE Early Career Research Program under award DE-SC0022924. AS is supported by a Mathematical Institute Studentship, Oxford and by the ERC grant GALOP ID: 724638. K.C. is supported by the NSERC Discovery Grant program and by the Perimeter Institute for Theoretical Physics. Research at Perimeter Institute is supported by the Government of Canada through Industry Canada and by the Province of Ontario through the Ministry of Research and Innovation.

\bibliographystyle{JHEP}
\bibliography{twist.bib}

\providecommand{\href}[2]{#2}\begingroup\raggedright\begin{thebibliography}{10}

\bibitem{BFSS}
T.~Banks, W.~Fischler, S.~H. Shenker, and L.~Susskind, {\it {M theory as a
  matrix model: A Conjecture}},  {\em Phys. Rev. D} {\bf 55} (1997) 5112--5128,
  [\href{http://arxiv.org/abs/hep-th/9610043}{{\tt hep-th/9610043}}].

\bibitem{Polchinski}
J.~Polchinski, {\it {M theory and the light cone}},  {\em Prog. Theor. Phys.
  Suppl.} {\bf 134} (1999) 158--170,
  [\href{http://arxiv.org/abs/hep-th/9903165}{{\tt hep-th/9903165}}].

\bibitem{Susskind}
L.~Susskind, {\it {Holography in the flat space limit}},  {\em AIP Conf. Proc.}
  {\bf 493} (1999), no.~1 98--112,
  [\href{http://arxiv.org/abs/hep-th/9901079}{{\tt hep-th/9901079}}].

\bibitem{BMN}
D.~E. Berenstein, J.~M. Maldacena, and H.~S. Nastase, {\it {Strings in flat
  space and pp waves from N=4 superYang-Mills}},  {\em JHEP} {\bf 04} (2002)
  013, [\href{http://arxiv.org/abs/hep-th/0202021}{{\tt hep-th/0202021}}].

\bibitem{IKKT}
N.~Ishibashi, H.~Kawai, Y.~Kitazawa, and A.~Tsuchiya, {\it {A Large N reduced
  model as superstring}},  {\em Nucl. Phys. B} {\bf 498} (1997) 467--491,
  [\href{http://arxiv.org/abs/hep-th/9612115}{{\tt hep-th/9612115}}].

\bibitem{CG}
K.~Costello and D.~Gaiotto, {\it {Twisted Holography}},
  \href{http://arxiv.org/abs/1812.09257}{{\tt arXiv:1812.09257}}.

\bibitem{CP1}
K.~Costello and N.~M. Paquette, {\it {Twisted Supergravity and Koszul Duality:
  A case study in AdS$_3$}},  {\em Commun. Math. Phys.} {\bf 384} (2021), no.~1
  279--339, [\href{http://arxiv.org/abs/2001.02177}{{\tt arXiv:2001.02177}}].

\bibitem{Note1}
These examples often work in the supergravity limit of the B-model, known as
  Kodaira-Spencer theory \cite {BCOV, CosLiBCOV}.

\bibitem{Costello:2017fbo}
K.~Costello, {\it {Holography and Koszul duality: the example of the $M2$
  brane}},  \href{http://arxiv.org/abs/1705.02500}{{\tt arXiv:1705.02500}}.

\bibitem{PW}
N.~M. Paquette and B.~R. Williams, {\it {Koszul duality in quantum field
  theory}},  \href{http://arxiv.org/abs/2110.10257}{{\tt arXiv:2110.10257}}.

\bibitem{Costello:2022wso}
K.~Costello and N.~M. Paquette, {\it {Celestial holography meets twisted
  holography: 4d amplitudes from chiral correlators}},
  \href{http://arxiv.org/abs/2201.02595}{{\tt arXiv:2201.02595}}.

\bibitem{Bu:2022dis}
W.~Bu and E.~Casali, {\it {The 4d/2d correspondence in twistor space and
  holomorphic Wilson lines}},  \href{http://arxiv.org/abs/2208.06334}{{\tt
  arXiv:2208.06334}}.

\bibitem{Costello:2022upu}
K.~Costello and N.~M. Paquette, {\it {On the associativity of one-loop
  corrections to the celestial OPE}},
  \href{http://arxiv.org/abs/2204.05301}{{\tt arXiv:2204.05301}}.

\bibitem{Note2}
See also \cite {Ball:2021tmb}, and see \cite {Ren:2022sws} for the highly
  stringent constraints associativity places on 4d EFT coefficients.

\bibitem{Costello:2021bah}
K.~J. Costello, {\it {Quantizing local holomorphic field theories on twistor
  space}},  \href{http://arxiv.org/abs/2111.08879}{{\tt arXiv:2111.08879}}.

\bibitem{Costello:2019jsy}
K.~Costello and S.~Li, {\it {Anomaly cancellation in the topological string}},
  {\em Adv. Theor. Math. Phys.} {\bf 24} (2020), no.~7 1723--1771,
  [\href{http://arxiv.org/abs/1905.09269}{{\tt arXiv:1905.09269}}].

\bibitem{Lebrun91explicitself-dual}
C.~LeBrun, {\it {Explicit self-dual metrics on $\CP^2\#\cdots\#\CP^2$}},  {\em
  J. Differential Geometry} {\bf 34} (1991) 223--253.

\bibitem{Strominger:2017zoo}
A.~Strominger, {\it {Lectures on the Infrared Structure of Gravity and Gauge
  Theory}},  \href{http://arxiv.org/abs/1703.05448}{{\tt arXiv:1703.05448}}.

\bibitem{Pasterski:2021raf}
S.~Pasterski, M.~Pate, and A.-M. Raclariu, {\it {Celestial Holography}},  in
  {\em {2022 Snowmass Summer Study}}, 11, 2021.
\newblock \href{http://arxiv.org/abs/2111.11392}{{\tt arXiv:2111.11392}}.

\bibitem{CPS}
K.~Costello, N.~M. Paquette, and A.~Sharma {\em {To appear}} (2022).

\bibitem{Mason:1991rf}
L.~J. Mason and N.~M.~J. Woodhouse, {\em {Integrability, self-duality, and
  twistor theory}}.
\newblock Oxford University Press, 1996.

\bibitem{burns1986twistors}
D.~Burns, {\it Twistors and harmonic maps},  {\em Amer. Math. Soc. conference
  talk, Charlotte, NC} (1986).

\bibitem{Dunajski:2019smi}
M.~Dunajski and P.~Tod, {\it {Conformal and isometric embeddings of
  gravitational instantons}},  \href{http://arxiv.org/abs/2001.00033}{{\tt
  arXiv:2001.00033}}.

\bibitem{lebrun1988counter}
C.~LeBrun, {\it {Counter-examples to the generalized positive action
  conjecture}},  {\em Commun. Math. Phys.} {\bf 118} (1988) 591--596.

\bibitem{lebrun1988poon}
C.~LeBrun, {\it {Poon's self-dual metrics and K\"ahler geometry}},  {\em J.
  differential geometry} {\bf 28} (1988) 341--343.

\bibitem{Hawking:1979pi}
S.~W. Hawking, D.~N. Page, and C.~N. Pope, {\it {Quantum Gravitational
  Bubbles}},  {\em Nucl. Phys. B} {\bf 170} (1980) 283--306.

\bibitem{Hartnoll:2004rv}
S.~A. Hartnoll and G.~Policastro, {\it {Spacetime foam in twistor string
  theory}},  {\em Adv. Theor. Math. Phys.} {\bf 10} (2006), no.~2 181--216,
  [\href{http://arxiv.org/abs/hep-th/0412044}{{\tt hep-th/0412044}}].

\bibitem{Note3}
The $\protect \text {K\"ahler}$ potential perturbation associated to a single
  closed-string field emanating from a D1 brane was computed in \cite
  {Costello:2021bah} to be $\protect \qopname \relax o{log}||u||^2$. What is
  surprising is that this already satisfies the equations of motion and that
  there are no further corrections.

\bibitem{Donaldson:1985zz}
S.~K. Donaldson, {\it {Anti self-dual Yang-Mills connections over complex
  algebraic surfaces and stable vector bundles}},  {\em Proc. Lond. Math. Soc.}
  {\bf 50} (1985) 1--26.

\bibitem{Nair:1991bf}
V.~P. Nair, {\it {Kahler-Chern-Simons theory}},
  \href{http://arxiv.org/abs/hep-th/9110042}{{\tt hep-th/9110042}}.

\bibitem{Losev:1995cr}
A.~Losev, G.~W. Moore, N.~Nekrasov, and S.~Shatashvili, {\it {Four-dimensional
  avatars of two-dimensional RCFT}},  {\em Nucl. Phys. B Proc. Suppl.} {\bf 46}
  (1996) 130--145, [\href{http://arxiv.org/abs/hep-th/9509151}{{\tt
  hep-th/9509151}}].

\bibitem{Note4}
This is the 4d analogue of the quantization of Kac-Moody level in 2d WZW
  models.

\bibitem{Yang:1977zf}
C.~N. Yang, {\it {Condition of Self-duality for SU(2) Gauge Fields on Euclidean
  Four-Dimensional Space}},  {\em Phys. Rev. Lett.} {\bf 38} (1977) 1377.

\bibitem{Ooguri:1991ie}
H.~Ooguri and C.~Vafa, {\it {N=2 heterotic strings}},  {\em Nucl. Phys. B} {\bf
  367} (1991) 83--104.

\bibitem{Bittleston:2020hfv}
R.~Bittleston and D.~Skinner, {\it {Twistors, the ASD Yang-Mills equations, and
  4d Chern-Simons theory}},  \href{http://arxiv.org/abs/2011.04638}{{\tt
  arXiv:2011.04638}}.

\bibitem{Penna:2020uky}
R.~F. Penna, {\it {Twistor Actions for Integrable Systems}},  {\em JHEP} {\bf
  09} (2021) 140, [\href{http://arxiv.org/abs/2011.05831}{{\tt
  arXiv:2011.05831}}].

\bibitem{Guevara:2021abz}
A.~Guevara, E.~Himwich, M.~Pate, and A.~Strominger, {\it {Holographic symmetry
  algebras for gauge theory and gravity}},  {\em JHEP} {\bf 11} (2021) 152,
  [\href{http://arxiv.org/abs/2103.03961}{{\tt arXiv:2103.03961}}].

\bibitem{Ball:2021tmb}
A.~Ball, S.~A. Narayanan, J.~Salzer, and A.~Strominger, {\it {Perturbatively
  exact w$_{1+\infty}$ asymptotic symmetry of quantum self-dual gravity}},
  {\em JHEP} {\bf 01} (2022) 114, [\href{http://arxiv.org/abs/2111.10392}{{\tt
  arXiv:2111.10392}}].

\bibitem{Fateev:1992tk}
V.~A. Fateev, E.~Onofri, and A.~B. Zamolodchikov, {\it {Integrable deformations
  of the $O(3)$ sigma model. The sausage model}},  {\em Nucl. Phys. B} {\bf
  406} (1993) 521--565.

\bibitem{Note5}
From the point of view of a 10d type I string compactified on twistor space,
  the fields arise from 3-7 and 3-3 strings. When placed in an $\Omega
  $-background transverse to twistor space, the D3 and D7-branes become
  Euclidean D1 and D5 branes.

\bibitem{Beem:2013sza}
C.~Beem, M.~Lemos, P.~Liendo, W.~Peelaers, L.~Rastelli, and B.~C. van Rees,
  {\it {Infinite Chiral Symmetry in Four Dimensions}},  {\em Commun. Math.
  Phys.} {\bf 336} (2015), no.~3 1359--1433,
  [\href{http://arxiv.org/abs/1312.5344}{{\tt arXiv:1312.5344}}].

\bibitem{SW}
N.~Seiberg and E.~Witten, {\it {Electric - magnetic duality, monopole
  condensation, and confinement in N=2 supersymmetric Yang-Mills theory}},
  {\em Nucl. Phys. B} {\bf 426} (1994) 19--52,
  [\href{http://arxiv.org/abs/hep-th/9407087}{{\tt hep-th/9407087}}]. [Erratum:
  Nucl.Phys.B 430, 485--486 (1994)].

\bibitem{MoosavianZhou}
R.~Lorgat, S.~F. Moosavian, and Y.~Zhao {\em {To appear}} (2022).

\bibitem{BRP}
C.~Beem, W.~Peelaers, and L.~Rastelli, {\it {unpublished work}}, .

\bibitem{Cordova:2017mhb}
C.~Cordova, D.~Gaiotto, and S.-H. Shao, {\it {Surface Defects and Chiral
  Algebras}},  {\em JHEP} {\bf 05} (2017) 140,
  [\href{http://arxiv.org/abs/1704.01955}{{\tt arXiv:1704.01955}}].

\bibitem{Arefeva:1974jv}
I.~Y. Arefeva, L.~D. Faddeev, and A.~A. Slavnov, {\it {Generating Functional
  for the S Matrix in Gauge Theories}},  {\em Teor. Mat. Fiz.} {\bf 21} (1974)
  311--321.

\bibitem{Jevicki:1987ax}
A.~Jevicki and C.-k. Lee, {\it {The S Matrix Generating Functional and
  Effective Action}},  {\em Phys. Rev. D} {\bf 37} (1988) 1485.

\bibitem{Rosly:1996vr}
A.~A. Rosly and K.~G. Selivanov, {\it {On amplitudes in selfdual sector of
  Yang-Mills theory}},  {\em Phys. Lett. B} {\bf 399} (1997) 135--140,
  [\href{http://arxiv.org/abs/hep-th/9611101}{{\tt hep-th/9611101}}].

\bibitem{Rosly:1997ap}
A.~A. Rosly and K.~G. Selivanov, {\it {Gravitational SD perturbiner}},
  \href{http://arxiv.org/abs/hep-th/9710196}{{\tt hep-th/9710196}}.

\bibitem{Hawking:1979hw}
S.~W. Hawking, D.~N. Page, and C.~N. Pope, {\it {The propagation of particles
  in space-time foam}},  {\em Phys. Lett. B} {\bf 86} (1979) 175--178.

\bibitem{BG}
K.~Budzik and D.~Gaiotto, {\it {Giant gravitons in twisted holography}},
  \href{http://arxiv.org/abs/2106.14859}{{\tt arXiv:2106.14859}}.

\bibitem{Atanasov:2021oyu}
A.~Atanasov, A.~Ball, W.~Melton, A.-M. Raclariu, and A.~Strominger, {\it {(2,
  2) Scattering and the celestial torus}},  {\em JHEP} {\bf 07} (2021) 083,
  [\href{http://arxiv.org/abs/2101.09591}{{\tt arXiv:2101.09591}}].

\bibitem{Crawley:2021auj}
E.~Crawley, A.~Guevara, N.~Miller, and A.~Strominger, {\it {Black Holes in
  Klein Space}},  \href{http://arxiv.org/abs/2112.03954}{{\tt
  arXiv:2112.03954}}.

\bibitem{Bittleston:2022nfr}
R.~Bittleston, A.~Sharma, and D.~Skinner, {\it {Quantizing the non-linear
  graviton}},  \href{http://arxiv.org/abs/2208.12701}{{\tt arXiv:2208.12701}}.

\bibitem{Adamo:2021bej}
T.~Adamo, L.~Mason, and A.~Sharma, {\it {Twistor sigma models for quaternionic
  geometry and graviton scattering}},
  \href{http://arxiv.org/abs/2103.16984}{{\tt arXiv:2103.16984}}.

\bibitem{Adamo:2021lrv}
T.~Adamo, L.~Mason, and A.~Sharma, {\it {Celestial $w_{1+\infty}$ Symmetries
  from Twistor Space}},  {\em SIGMA} {\bf 18} (2022) 016,
  [\href{http://arxiv.org/abs/2110.06066}{{\tt arXiv:2110.06066}}].

\bibitem{Adamo:2022mev}
T.~Adamo, L.~Mason, and A.~Sharma, {\it {Graviton scattering in self-dual
  radiative space-times}},  \href{http://arxiv.org/abs/2203.02238}{{\tt
  arXiv:2203.02238}}.

\bibitem{Dunajski:2006vs}
M.~Dunajski and S.~A. Hartnoll, {\it {Einstein-Maxwell gravitational instantons
  and five dimensional solitonic strings}},  {\em Class. Quant. Grav.} {\bf 24}
  (2007) 1841--1862, [\href{http://arxiv.org/abs/hep-th/0610261}{{\tt
  hep-th/0610261}}].

\bibitem{Pasterski:2020pdk}
S.~Pasterski and A.~Puhm, {\it {Shifting spin on the celestial sphere}},  {\em
  Phys. Rev. D} {\bf 104} (2021), no.~8 086020,
  [\href{http://arxiv.org/abs/2012.15694}{{\tt arXiv:2012.15694}}].

\bibitem{Fan:2022vbz}
W.~Fan, A.~Fotopoulos, S.~Stieberger, T.~R. Taylor, and B.~Zhu, {\it {Elements
  of celestial conformal field theory}},  {\em JHEP} {\bf 08} (2022) 213,
  [\href{http://arxiv.org/abs/2202.08288}{{\tt arXiv:2202.08288}}].

\bibitem{Casali:2022fro}
E.~Casali, W.~Melton, and A.~Strominger, {\it {Celestial Amplitudes as
  AdS-Witten Diagrams}},  \href{http://arxiv.org/abs/2204.10249}{{\tt
  arXiv:2204.10249}}.

\bibitem{Gonzo:2022tjm}
R.~Gonzo, T.~McLoughlin, and A.~Puhm, {\it {Celestial holography on Kerr-Schild
  backgrounds}},  \href{http://arxiv.org/abs/2207.13719}{{\tt
  arXiv:2207.13719}}.

\bibitem{BCOV}
M.~Bershadsky, S.~Cecotti, H.~Ooguri, and C.~Vafa, {\it {Kodaira-Spencer theory
  of gravity and exact results for quantum string amplitudes}},  {\em Commun.
  Math. Phys.} {\bf 165} (1994) 311--428,
  [\href{http://arxiv.org/abs/hep-th/9309140}{{\tt hep-th/9309140}}].

\bibitem{CosLiBCOV}
K.~J. Costello and S.~Li, {\it {Quantum BCOV theory on Calabi-Yau manifolds and
  the higher genus B-model}},  \href{http://arxiv.org/abs/1201.4501}{{\tt
  arXiv:1201.4501}}.

\bibitem{Ren:2022sws}
L.~Ren, M.~Spradlin, A.~Yelleshpur~Srikant, and A.~Volovich, {\it {On Effective
  Field Theories with Celestial Duals}},
  \href{http://arxiv.org/abs/2206.08322}{{\tt arXiv:2206.08322}}.

\end{thebibliography}\endgroup


\onecolumngrid 

\newpage

\begin{center}
    \bf\large{Supplementary Material}
\end{center}

\medskip

In this supplement, we discuss some subtleties in deriving OPEs in the holographic dual chiral algebra using Wick contractions. Following this, we return to the bulk and provide sample computations of two- and three-point amplitudes on Burns space. Specifically, we show that the resulting amplitudes are what one would expect from the OPE of leading soft gluon operators $j_a(z) \equiv J_a[0,0](z)$,
\be\label{jjope}
j_a(z_1)\,j_b(z_2) \sim -\frac{2N\kappa_{ab}}{z_{12}^2} + \frac{f_{ab}{}^cj_c(z_2)}{z_{12}}\,,
\ee
that one finds from Wick contracting the operators $j_{ij}(z) = I_iI_j(z)$ of the dual chiral algebra. Here, $a$ is an adjoint index of $\mathfrak{so}(8)$ which translates to $ij$ in double-line notation. It is notable that one can naturally reproduce the negative level $-2N$ of the chiral algebra from the bulk. More details of such computations will appear in the companion paper.

\section{Chiral algebra computations}

Recall that the dual chiral algebra is an $\Sp(N)$ gauge theory. Applying the BRST procedure for the $\Sp(N)$ gauge group generates the constraint equation 
\begin{equation}
    [X_{\dot{1}}, X_{\dot{2}} ]   + \delta^{ij} I_i I_j = 0\,.
\end{equation}
This is the familiar ADHM equation for $\SO(8)$ instantons on $\R^4$.  It tells us that we can commute $X_{\dot{1}}$ and $X_{\dot{2}}$ in the strings $I_i X_{\dot{1}}^n X_{\dot{2}}^m I_j$ at the price of introducing a double-trace operator, i.e. an operator which is a product of two open string states.

In the planar limit, OPEs which take two open string states to a product of several open string states are suppressed.  This means that when we are computing planar OPEs of open string states, we can treat the matrices $X_{\dot{1}}, X_{\dot{2}}$ as commuting.

Writing $[X\,i] = \eps^{\dal\dot\beta}\tilde{\lambda}_{i\dot \alpha} X_{\dot \beta}$, we can compute the OPEs 
\begin{equation}
    \frac{(2N)^{-\frac{m}{2}}}{m!} I_i [X\,1]^m I_j (z_1)\; \frac{(2N)^{-\frac{n}{2}}}{n!} I_k [X\,2]^n I_l(z_2) 
\end{equation}
by Wick contraction of adjacent fields in each word.  Contracting only adjacent fields is forced by working in the planar limit.  We must contract the last $I$ in the first word with the first $I$ in the second word, then the last $[X\,1]$ in the first word with the first $[X\,2]$ in  the second word, and so on.  Due to the antisymmetry $I_k [X\,2]^n I_l = - I_l [X\,2]^n I_k$,
contractions could also be done in the opposite order, by contracting the last $I$ in the second word with the last $I$ in the first word, the last $[X\,1]$ with the last $[X\,2]$, and so on. Then the planar OPE reads 
\be\label{opeblock}
\begin{split}
    \frac{(2N)^{-\frac{m}{2}}}{m!} I_i [X\,1]^m I_j (z_1)\; \frac{(2N)^{-\frac{n}{2}}}{n!} I_k [X\,2]^n I_l(z_2)  &\sim -\frac{2N\,(\delta_{il}\delta_{jk}-\delta_{ik}\delta_{jl}) \,\delta_{mn}\mathbf{1}}{m!\,n!}\, \frac{  \,(-[1\,2])^n}{z_{12}^{n+2}}\\
    &\hspace{3cm}+ \text{terms with fewer contractions}\,.
\end{split}
\ee
We have only displayed terms that contribute to the identity block in the OPE and the corresponding two-point function.    

It is important to emphasize that this is the planar OPE.  The full OPE will contain terms on the right hand side that are subleading in $N$. For example, if $m = n = 1$, an exact evaluation of equation \eqref{opeblock} will involve the dimension of the representation $\C^2 \otimes \bigwedge^2_\text{trace-free} \C^{2 N}$ containing $X$. This is $(2N)^2 - 2N - 2$.  In the planar limit, only the dominant term $(2 N)^2$ of this expression plays a role, and this becomes $2N$ after accounting for the factors of $2N$ on the left hand side.   The exact evaluation of \eqref{opeblock} would involve replacing $2 N$ on the right hand side by $2 N-1-1/2N$. 

\section{Two-point function}

We study the scattering of states labeled by complex null momenta $p_{i\al\dal} = \lambda_{i\al}\tilde\lambda_{i\dal}$, with $\lambda_{i\al} = (z_i^{-1},1)$ and $i$ being the particle label. Their wavefunctions can be expanded as a power series,
\be\label{phiseries}
\phi_{i} = \mathsf{T}_{a_i}\sum_{n=0}^\infty\phi_i^{(n)}\,,\qquad\phi_i^{(n)} = \frac{z_i^{-1}}{(2n)!}\left(-\frac{1}{z_i}\frac{[u\,i][\hat u\,i]}{||u||^2}\right)^n{}_1F_1(n+1,2n+1\,|\,\im p_i\cdot x)
\ee
where recall that $u^{\dal}=x^{1\dal},\hat u^{\dal}=x^{2\dal}$ and $||u||^2=x^2/2\equiv\frac{1}{2}\delta_{\mu\nu}x^\mu x^\nu$ (not to be confused with $x^{2\dal}$). In particular, $p_i\cdot x = z_i^{-1}[u\,i]+[\hat u\,i]$. We also make use of the standard integral representation
\be\label{intrep}
\frac{1}{(2n)!}\,{}_1F_1(n+1,2n+1\,|\,\im p_i\cdot x) = \frac{1}{n!\,(n-1)!}\int_0^1\d s\;s^{n}\,(1-s)^{n-1}\,\e^{\im s p_i\cdot x}\,,\qquad n\geq1
\ee
of the confluent hypergeometric functions that occur in $\phi_i^{(n)}$. For $n=0$, we instead simply find ${}_1F_1(1,1\,|\,\im p_i\cdot x)=\e^{\im p_i\cdot x}$.

The 2-point amplitude of momentum eigenstates vanishes in flat space but gives a non-trivial result on Burns space. Starting with the on-shell kinetic term,
\be
A(1,2) = \frac{N}{8\pi^2}\int_{\widetilde\C^2}\p\dbar K\wedge\tr\left(\p\phi_{1}\wedge\dbar\phi_{2}+\p\phi_{2}\wedge\dbar\phi_{1}\right)
\ee
and plugging in \eqref{phiseries} along with $K=||u||^2+\log||u||^2$, one can write this as an infinite sum, each containing two parts:
\be\label{A12exp}
A(1,2) = \tr(\msf{T}_{a_1}\msf{T}_{a_2})\sum_{n=0}^\infty\bigl( M_n(1,2) + N_n(1,2)\bigr)\,.
\ee
The two summands are given by
\begin{align}
    M_n(1,2) &= \frac{N}{8\pi^2}\,\int_{\widetilde\C^2}\frac{u^{\dal}\d u_{\dal}\wedge\hat u^{\dot\beta}\d \hat u_{\dot\beta}}{||u||^4}\wedge\left(\sum_{k+l=n}\p\phi_1^{(k)}\wedge\dbar\phi_2^{(l)} + (1\leftrightarrow2)\right)\,,\\
    N_n(1,2) &= \frac{N}{8\pi^2}\int_{\widetilde\C^2}\d u_{\dal}\wedge\d \hat u^{\dal}\wedge\left(\sum_{k+l=n+1}\p\phi_1^{(k)}\wedge\dbar\phi_2^{(l)} + (1\leftrightarrow2)\right)\,,
\end{align}
where spinor indices are raised or lowered using the Levi-Civita symbol, e.g., $\d u^{\dal} = \eps^{\dal\dot\beta}\d u_{\dot\beta}$, $\d u_{\dal} = \d u^{\dot\beta}\eps_{\dot\beta\dal}$, etc. In writing this, we have dropped the trivial flat space contribution
\be
\frac{N}{8\pi^2}\int_{\widetilde\C^2}\d u_{\dal}\wedge\d \hat u^{\dal}\wedge\left(\p\phi_1^{(0)}\wedge\dbar\phi_2^{(0)} + (1\leftrightarrow2)\right) \propto (p_1+p_2)^2\;\delta^4(p_1+p_2) = 0
\ee
from the expansion \eqref{A12exp}.

Under a simultaneous rescaling $\tilde\lambda_i\mapsto t\,\tilde\lambda_i$ for $i=1,2$, we can rescale the integration variable $x^{\al\dal}\mapsto t^{-1}x^{\al\dal}$ to show that $M_n$, $N_n$ scale as $M_n\mapsto t^{2n}M_n$, $N_n\mapsto t^{2n}N_n$. Combined with the invariance of the analytically continued amplitudes under $\SL(2,\C)$ rotations of the dotted index on $\tilde\lambda_i^{\dal}$, we conclude that $M_n,N_n\propto[1\,2]^{n}$. As a result, $M_{n}+N_{n}$ precisely contributes to 2-point functions of the currents $J_a[k,l]$ for $k+l=n$. In particular, $M_0+N_0$ is expected to give rise to the 2-point function of $J_a[0,0]$.

Let us compute $M_0$, $N_0$. Plugging in $\phi_i^{(0)} = z_i^{-1}\,\e^{\im p_i\cdot x}$ reduces $M_0$ to
\be\label{M0eq}
\begin{split}
    M_0(1,2) &= \frac{N}{8\pi^2}\,\int_{\widetilde\C^2}\frac{u^{\dal}\d u_{\dal}\wedge\hat u^{\dot\beta}\d \hat u_{\dot\beta}}{||u||^4}\wedge\left(\p\phi_1^{(0)}\wedge\dbar\phi_2^{(0)} + (1\leftrightarrow2)\right)\\
    &=\frac{N}{2\pi^2}\int_{\R^4}\d^4x\;\e^{\im(p_1+p_2)\cdot x}\;\frac{x^{\al\dal}x^{\beta\dot\beta}}{(x^2)^2}\;\frac{o_\al\iota_\beta(z_2\tilde\lambda_{1\dal}\tilde\lambda_{2\dot\beta} + z_1\tilde\lambda_{2\dal}\tilde\lambda_{1\dot\beta})}{(z_1z_2)^2}
\end{split}
\ee
where $o_\al=(1,0)$, $\iota_\al=(0,1)$ and we have plugged in $u^{\dal}=x^{\al\dal}o_\al$, $\hat u^{\dal}=x^{\al\dal}\iota_\al$. To compute this, we need the Fourier integral
\be\label{logfourier}
\int\d^4x\;\e^{\im k\cdot x}\;\frac{x^{\al\dal} x^{\beta\dot\beta}}{x^4} = \frac{2\pi^2}{k^2}\,\biggl(\eps^{\al\beta}\eps^{\dal\dot\beta} - \frac{2\,k^{\al\dal} k^{\beta\dot\beta}}{k^2}\biggr)\,,
\ee
which we analytically continue to complex $k_\mu$ away from its singular locus $k^2\equiv\delta^{\mu\nu}k_\mu k_\nu=0$. Inserting this into \eqref{M0eq} for the value $k=p_1+p_2$ results in
\be
M_0(1,2) = \frac{N}{(p_1+p_2)^2}\,\biggl(\eps^{\al\beta}\eps^{\dal\dot\beta} - \frac{2\,(p_1+p_2)^{\al\dal} (p_1+p_2)^{\beta\dot\beta}}{(p_1+p_2)^2}\biggr)\,\frac{o_\al\iota_\beta(z_2\tilde\lambda_{1\dal}\tilde\lambda_{2\dot\beta} + z_1\tilde\lambda_{2\dal}\tilde\lambda_{1\dot\beta})}{(z_1z_2)^2}\,.
\ee
Using $(p_1+p_2)^{\al\dal} = \lambda_1^\al\tilde\lambda_1^{\dal}+\lambda_2^\al\tilde\lambda_2^{\dal}$ along with $(p_1+p_2)^2 = 2\,p_1\cdot p_2 = 2\,z_{12}[1\,2]/z_1z_2$, this promptly collapses to
\be
M_0(1,2) = -\frac{N}{z_{12}^2}\,.
\ee
In computing this, notice how the factors of $[1\,2]$ cancel out, corroborating our prediction for the powers of $[1\,2]$.

Next, we need to calculate the second summand
\be\label{N012}
N_0(1,2) = \frac{N}{8\pi^2}\int_{\widetilde\C^2}\d u_{\dal}\wedge\d \hat u^{\dal}\wedge\left[\p\phi_1^{(1)}\wedge\dbar\phi_2^{(0)} + \p\phi_1^{(0)}\wedge\dbar\phi_2^{(1)} +(1\leftrightarrow2)\right]\,.
\ee
Using the integral representation \eqref{intrep}, the first term in this expression yields
\be
\frac{N}{8\pi^2}\int_{\widetilde\C^2}\d u_{\dal}\wedge\d \hat u^{\dal}\wedge\p\phi_{1}^{(1)}\wedge\dbar\phi_{2}^{(0)} =\frac{N}{4\pi^2}\int\d^4x\int_0^1\d s\,s\;\frac{\e^{\im(sp_1+p_2)\cdot x}}{z_1^2z_2}\left(\frac{s\,[1\,2]\la o|x|1]\la\iota|x|1]}{z_1\,x^2}-\frac{2\,\im\,\la o|x|2]\la\iota|x|1]^2}{x^4}\right)\,,
\ee
having introduced the notation $\la\xi|x|\tilde\lambda] = x^{\al\dal}\xi_\al\tilde\lambda_{\dal}$. Applying the analytically continued Fourier integrals
\begin{align}
    \int\d^4x\;\e^{\im k\cdot x}\;\frac{x^{\al\dal} x^{\beta\dot\beta}}{x^2} &= \frac{8\pi^2}{k^4}\,\biggl(\eps^{\al\beta}\eps^{\dal\dot\beta}-\frac{4\,k^{\al\dal} k^{\beta\dot\beta}}{k^2}\biggr)\,,\label{4id1}\\
    \int\d^4x\;\e^{\im k\cdot x}\;\frac{x^{\al\dal} x^{\beta\dot\beta} x^{\gamma\dot\gamma}}{x^4} &= \frac{4\pi^2\im}{k^4}\,\biggl(\eps^{\al\beta}\eps^{\dal\dot\beta}k^{\gamma\dot\gamma} + \eps^{\beta\gamma}\eps^{\dot\beta\dot\gamma}k^{\al\dal}+\eps^{\gamma\al}\eps^{\dot\gamma\dal}k^{\beta\dot\beta}-\frac{4\,k^{\al\dal} k^{\beta\dot\beta} k^{\gamma\dot\gamma}}{k^2}\biggr)\label{4id2}
\end{align}
for the value $k=sp_1+p_2$, one can check that this vanishes identically:
\be
\frac{N}{8\pi^2}\int_{\widetilde\C^2}\d u_{\dal}\wedge\d \hat u^{\dal}\wedge\p\phi_{1}^{(1)}\wedge\dbar\phi_{2}^{(0)} = \frac{N}{4\pi^2}\int_0^1\frac{\d s}{s}\left[\frac{4\pi^2 z_2}{z_{12}^3}+\frac{4\pi^2 }{z_{12}^2}\left(1-\frac{z_1}{z_{12}}\right)\right]=0\,.
\ee
Similarly, we can verify that the other terms in \eqref{N012} also vanish, leading to
\be
N_0(1,2) = 0\,.
\ee
Normalizing Lie algebra generators such that $\tr(\msf{T}_{a_1}\msf{T}_{a_2}) = 2\,\kappa_{a_1a_2}$, we finally find
\be
A(1,2) = -\frac{2N\kappa_{a_1a_2}}{z_{12}^2} + O([1\,2])\,.
\ee
This $O([1\,2]^0)$ term is indeed the expected 2-point function of the leading soft gluon current $j_a\equiv J_a[0,0]$ in the dual defect CFT (up to the usual conformal block ambiguity mentioned in the text). The same techniques can be scaled up to compute the full 2-point amplitude given in \eqref{2ptamp}, as we will describe in the companion paper.

As we remarked above, on the chiral algebra side, an exact computation of the two-point function is readily obtainable, including $1/N$ corrections.  These corrections should correspond to loop contributions to the two-point function of the $\WZW_4$ model. It would be very interesting to explicitly compute such loop amplitudes and match them with non-planar corrections in the chiral algebra. 

\section{Three-point function}

To really test our holographic proposal, we need to compute the 3-gluon amplitude from the interaction vertex
\be
-\frac{N}{24\pi^2}\int_{\widetilde\C^2}\p\dbar K\wedge\tr\,\phi\,[\p\phi,\dbar\phi]\,.
\ee
To do this, substitute $\phi=\sum_{i=1}^3\varepsilon_i\phi_i$ into this vertex. The 3-point amplitude is then the coefficient of $\varepsilon_1\varepsilon_2\varepsilon_3$:
\be
A(1,2,3) = -\frac{N}{24\pi^2}\int\p\dbar K\wedge\tr\left\{\phi_3\left([\p\phi_1,\dbar\phi_2]+[\p\phi_2,\dbar\phi_1]\right)+\text{cyclic}\right\}\,.
\ee
Substituting $K=||u||^2+\log||u||^2$ and $\phi_i=\sum_{n\geq0}\phi_i^{(n)}$, this breaks into
\be\label{A123}
A(1,2,3) = f_{a_1a_2a_3}\sum_{n=0}^\infty\left(M_n(1,2,3) + N_n(1,2,3)\right)
\ee
having defined the totally antisymmetric structure constants $f_{a_1a_2a_3} = f_{a_1a_2}{}^c\kappa_{ca_3}$. The two summands read
\begin{align}
    M_n(1,2,3) &= -\frac{N}{12\pi^2}\int\frac{u^{\dal}\d u_{\dal}\wedge\hat u^{\dot\beta}\d \hat u_{\dot\beta}}{||u||^4}\wedge\sum_{k+l+m=n}\phi_{3}^{(m)}\left(\p\phi_{1}^{(k)}\wedge\dbar\phi_{2}^{(l)}-\p\phi_{2}^{(l)}\wedge\dbar\phi_{1}^{(k)}\right)+\text{cyclic}\,,\label{An0123}\\
    N_n(1,2,3) &= -\frac{N}{12\pi^2}\int\d u_{\dal}\wedge\d \hat u^{\dal}\wedge\sum_{k+l+m=n+1}\phi_{3}^{(m)}\left(\p\phi_{1}^{(k)}\wedge\dbar\phi_{2}^{(l)}-\p\phi_{2}^{(l)}\wedge\dbar\phi_{1}^{(k)}\right)+\text{cyclic}\,.\label{An1123}
\end{align}
Under the simultaneous rescaling $\tilde\lambda_i\mapsto t\,\tilde\lambda_i$ for $i=1,2,3$, $M_n\mapsto t^{2n} M_n$, $N_n\mapsto t^{2n}N_n$. Therefore, $M_n,N_n$ will be linear combinations of monomials $[1\,2]^k[2\,3]^l[3\,1]^m$ with $k+l+m=n$ and so contribute precisely to the three-point function of three currents $J_a[i_1,j_1]$, $J_b[i_2,j_2]$, $J_c[i_3,j_3]$ such that $\sum_{\ell=1}^3(i_\ell+j_\ell)=2n$.

In writing \eqref{A123}, we have dropped the distributional contribution
\begin{multline}
-\frac{N}{12\pi^2}\int\d u_{\dal}\wedge\d \hat u^{\dal}\wedge\phi_{3}^{(0)}\left(\p\phi_{1}^{(0)}\wedge\dbar\phi_{2}^{(0)}-\p\phi_{2}^{(0)}\wedge\dbar\phi_{1}^{(0)}\right)+\text{cyclic}\\
= 8\pi^2N\,[1\,2]\,\delta(z_{12})\,\delta(z_{23})\,\delta([1\,2]-[2\,3])\,\delta([2\,3]-[3\,1])    
\end{multline}
of the flat space 3-point tree amplitude. Due to momentum conservation, this takes the form of a contact term with support at $z_1=z_2=z_3$ and $[1\,2]=[2\,3]=[3\,1]$. We do not know how to reproduce such contact terms from the chiral algebra OPE, so we will work at general separations $z_{ij}\neq0$. In fact, the 3-point amplitude is the only outlier, since all other flat space amplitudes of $\SO(8)$ WZW$_4$ coupled to scalar-flat K\"ahler gravity are expected to vanish due to its quantum integrability. This is because such an integrable theory has an infinite dimensional symmetry algebra whose generators have non-trivial Lorentz charge, so it is prohibited from exhibiting 4- and higher point amplitudes by the Coleman-Mandula theorem. It can only exhibit non-trivial amplitudes in \emph{curved} spacetimes like Burns space.

Next let us compute the leading piece $M_0(1,2,3)$,
\be
M_0(1,2,3) = -\frac{N}{12\pi^2}\int\frac{u^{\dal}\d u_{\dal}\wedge\hat u^{\dot\beta}\d \hat u_{\dot\beta}}{||u||^4}\wedge\left\{\phi_{3}^{(0)}\left(\p\phi_{1}^{(0)}\wedge\dbar\phi_{2}^{(0)}-\p\phi_{2}^{(0)}\wedge\dbar\phi_{1}^{(0)}\right)+\text{cyclic}\right\}\,.
\ee
Substituting $\phi_{i}^{(0)}=z_i^{-1}\e^{\im p_i\cdot x}$ yields
\be
    M_0(1,2,3) = -\frac{N}{3\pi^2}\int\d^4x\;\frac{\e^{\im(p_1+p_2+p_3)\cdot x}}{z_1z_2z_3}\left(\frac{z_2\la o|x|1]\la\iota|x|2]-z_1\la o|x|2]\la\iota|x|1]}{z_1z_2\,x^4}+\text{cyclic}\right)\,.
\ee
The Fourier integral \eqref{logfourier} evaluated at $k=p_1+p_2+p_3$ reduces this to
\be\label{A10123}
M_0(1,2,3) = \frac{4N}{(z_1z_2z_3)^2}\,\frac{[1\,2]\,(z_2[1\,3]+z_1[2\,3])}{(p_1+p_2+p_3)^4}+\text{cyclic}\,.
\ee
Here, $(p_1+p_2+p_3)^2 = 2\,(z_3z_{12}[1\,2]+z_1z_{23}[2\,3]+z_2z_{31}[3\,1])/z_1z_2z_3$ in our parametrization $\lambda_{i\al} = (z_i^{-1},1)$.

The second piece at this order is
\begin{multline}
N_0(1,2,3) = -\frac{N}{12\pi^2}\int\d u_{\dal}\wedge\d\hat u^{\dal}\wedge\Bigl[\phi_{3}^{(0)}\left(\p\phi_{1}^{(1)}\wedge\dbar\phi_{2}^{(0)}-\p\phi_{2}^{(0)}\wedge\dbar\phi_{1}^{(1)}\right)+ \phi_{3}^{(0)}\left(\p\phi_{1}^{(0)}\wedge\dbar\phi_{2}^{(1)}-\p\phi_{2}^{(1)}\wedge\dbar\phi_{1}^{(0)}\right) \\
+ \phi_{3}^{(1)}\left(\p\phi_{1}^{(0)}\wedge\dbar\phi_{2}^{(0)}-\p\phi_{2}^{(0)}\wedge\dbar\phi_{1}^{(0)}\right)
+\text{cyclic}\Bigr]\,.
\end{multline}
Using the integral representation \eqref{intrep} and Fourier identities like \eqref{4id1}, \eqref{4id2}, we can reduce this to
\be\label{A11123}
N_0(1,2,3) = -\frac{4N}{(z_1z_2z_3)^2}\,\frac{z_1+z_2}{z_{12}}\,\frac{([1\,3]+[2\,3])(z_2[1\,3]+z_1[2\,3])}{(p_1+p_2+p_3)^4}+\text{cyclic}\,.
\ee
Adding \eqref{A11123} to \eqref{A10123} leads to a dramatic simplification
\be
M_0(1,2,3) + N_0(1,2,3) =  -\frac{2N}{z_{12}z_{13}z_{23}}\,,
\ee
yielding the 3-point amplitude
\be
A(1,2,3) = -\frac{2Nf_{a_1a_2a_3}}{z_{12}z_{13}z_{23}} + O([1\,2],[2\,3],[3\,1])\,.
\ee
This agrees with the expectation from the $j_a(z_1)\,j_b(z_2)$ OPE \eqref{jjope} of a current algebra at level $-2N$. 

\end{document}